\documentclass[aps,prl,amsmath,amssymb,floatfix,lengthcheck,reprint]{revtex4-1}

\usepackage{graphicx}

\begin{document}
\title{Noise Performance of Lumped Element Direct Current Superconducting Quantum Interference Device Amplifiers in the 4 GHz-8 GHz Range\footnote{This paper is a contribution of the U.S. government and is not subject to U.S. copyright.}}
\author{Lafe Spietz}
 \email{lafe.spietz@boulder.nist.gov}
\author{Kent Irwin}
\author{Minhyea Lee}
\author{Jos\'e Aumentado}%
\affiliation{National Institute of Standards and Technology, Boulder, Colorado 80305, USA}

\date{\today}

\begin{abstract}
	We report on the noise of a lumped element Direct Current Superconducting Quantum Interference Device amplifier.  We show that the noise temperature in the 4 GHz-8 GHz range over ranges of 10's of MHz is below 1 kelvin (three photons of added noise), characterize the overall behavior of the noise as a function of bias parameters, and discuss potential mechanisms which determine the noise performance in this amplifier.  We show that this device can provide more than a factor of 10 improvement in practical system noise over existing phase-preserving microwave measurement systems in this frequency band.
\end{abstract}

\maketitle

	The measurement needs of the quantum information community have created a rapidly growing demand for ultra-low-noise microwave amplifiers.  The typical frequencies of interest for quantum information are in the 4 GHz--8 GHz range.  In general, the first stage cryogenic amplifier sets the overall speed of the measurement, which can be critical for quantum information experiments.  The commercial state of the art in cryogenic microwave amplifiers uses High Electron Mobility Transistor (HEMT) technology, and typically have a noise temperature of 2-4 K, but yield practical system noise temperatures in the 10-20 K range, as discussed below.  A rapidly growing subfield has emerged using superconducting electronics to build pre-amplifiers for these semiconductor-based cryogenic amplifiers.

	In the 4 GHz--8 GHz range, this has primarily consisted of parametric amplifiers.  Parametric amplifiers have demonstrated noise below the standard quantum limit and to have sufficient gain to overcome the following stages of amplifier noise \cite{manuel_2008}.  However, they have practical bandwidth typically well below 1 MHz and do not amplify both phase quadratures of a signal.  Furthermore, parametric amplifiers require a microwave bias, increasing significantly the complexity of operation.  Amplifiers using Superconducting Quantum Interference Devices (SQUIDs) have been shown to have near-quantum-limited noise performance in a variety of circumstances \cite{mueck_quantum}, primarily at frequencies below 2 GHz, or with limited gain \cite{prokopenko1}.  We have previously shown over 27 GHz of gain-bandwidth product, and gains up to 30 dB in microwave amplifiers using SQUIDs \cite{spietz_squidamp2}, and have characterized the input impedance of these devices \cite{spietz_squidamp1}.  Here we show noise performance in a SQUID amplifier with less than three photons per mode of added noise over 10's of MHz of instantaneous bandwidth in the 4 GHz--8 GHz range.
		
	Before discussing the SQUID amplifier in detail, we address the distinction between system noise and amplifier noise in practical microwave measurements.  When loss of a linear factor A at temperature T is placed between a signal and an amplifier of noise temperature $T_n^{amp}$, the combination of the amplifier and the loss acts like an amplifier with lower gain and higher noise.  The system noise temperature is 	
\begin{equation}
T_n^{sys} = AT_n^{amp} + (A-1)T,
\end{equation}
and the gain is suppressed by the factor A \cite{pozar}.  Because HEMT amplifiers dissipate 10's to 100's of milliwatts of power, they must be at the 4 K stage in a dilution refrigerator instead of at the base temperature due to cooling power considerations.  This forces the amplifier to be physically separated from the experiment by 10's of cm of cable, as well as by components such as isolators and couplers.  This in turn forces the noise temperature observed in typical experiments to be considerably higher than the intrinsic noise of the amplifiers.  Since the integration time required for a given signal to noise ratio scales with the square of the noise temperature \cite{dicke}, the factor of 10 practical system noise improvements shown with this amplifier can lead to as much as a factor of 100 shorter measurement time, making dramatic improvements in the types of experiments which are accessible with microwaves at very low temperatures.
		
\begin{figure}
\includegraphics[width=\columnwidth]{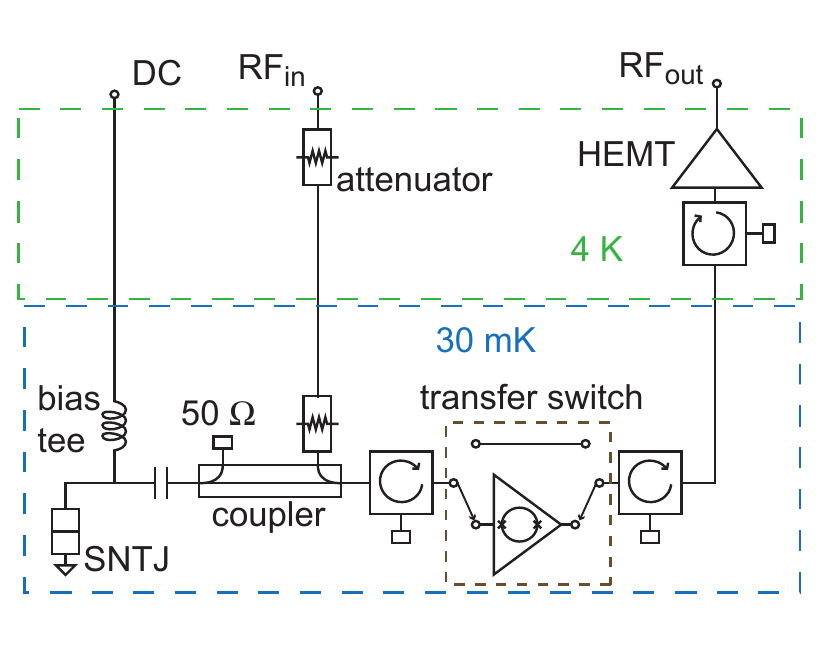}
\caption{Experimental layout.  This shows the apparatus with the SNTJ.  For the y-factor measurements, the apparatus before the directional coupler is replaced by the heated microwave termination.  The transfer switch connected to amplifier by superconducting cables allows for calibrated gain measurements.  Use of double isolators at both the input and output of the SQUID amplifier realistically simulates a typical quantum measurement experiment.}
\end{figure}					
				
	The devices measured here were lumped element DC-SQUID amplifiers, impedance matched at the input with quarter wave resonators of varying length to set the frequency, and impedance matched to 50 $\Omega$ at the output using series inductors and capacitors to ground, all on-chip, as described in \cite{spietz_squidamp1,spietz_squidamp2}.
				
	We have measured the system noise of our SQUID amplifiers in two different ways: the y-factor method and using a Shot Noise Tunnel Junction (SNTJ)\cite{inprep}.  Characterization of input impedance, bandwidth, and gain of these amplifiers is reported in previous work \cite{spietz_squidamp1,spietz_squidamp2}.  Figure 1 shows the apparatus used in this experiment, which was intended to realistically replicate the circumstances of a typical cryogenic microwave measurement, including the double isolator between the input of the SQUID amplifier and the noise source.  The y-factor method is the more conceptually straightforward and widely accepted, but the SNTJ method is much faster, and is thus much more useful for acquiring data into the rich structure of the noise as a function of the two bias parameters: the flux bias and the current bias. We used the y-factor measurements as well as shot noise to characterize the 7.06 GHz amplifier shown in Figure 2 as well as the 1.7 GHz amplifier, and just shot noise to characterize amplifiers at 6.3, 6.8, and 7.2 GHz.

	The SNTJ consists of a normal metal tunnel junction matched to 50 $\Omega$ over a broad band, and biased through a bias tee so that a known direct current can be applied to the junction while the microwave noise is measured by the amplifier \cite{shotnoisescience}.  Because the physics of this device is well understood, this measurement allows for the rapid determination of both the physical temperature of the sample and the noise temperature of the amplifier.  A detailed description of the use of the SNTJ for amplifier calibration will be given elsewhere.  The most important aspect of it for the purpose of this paper is its speed and ease of use relative to the y-factor method.  Our y-factor measurements were carried out by placing a heater, a ruthenium oxide thermometer and a microwave terminator in thermal contact with each other at the end of approximately 10 cm of niobium semi-rigid coaxial cable with a copper wire providing a known thermal conductance to the base temperature of the dilution refrigerator.  Changing this temperature by enough to get useful data on noise temperature took well over an hour for each thermal cycle.  The SNTJ, in contrast, can be switched in well under a millisecond, allowing for noise temperature measurements limited primarily by the speed of the microwave noise measurement, which is generally on the order of a second for a single trace from the spectrum analyzer.

\begin{figure}
\includegraphics{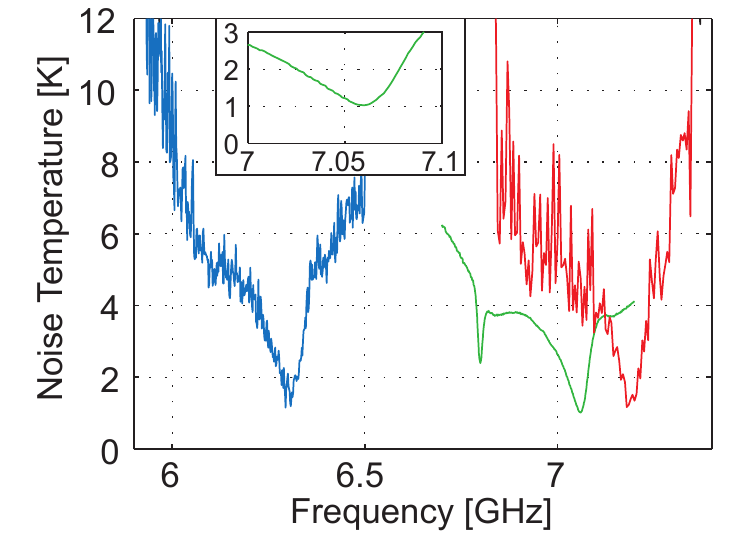}
\caption{System noise temperature as a function of frequency at a good bias point for three amplifiers with different lengths of input resonator.  The blue line is a 2 mm long input resonator, the green line is 1.6 mm, and the red line is 1.4 mm. Inset shows zoomed in noise temperature as a function of frequency around the optimal noise point for the 1.6 mm amplifier.  These data were taken using the y-factor method for the 1.6 mm device and the SNTJ method for the other devices, with the configuration of double circulators, directional coupler, and transfer switch shown in Figure 1.}
\end{figure}

	Figure 2 shows noise temperature data for three of our amplifiers at an optimal bias point, as measured using the y-factor method and the SNTJ method.  Typical system noise temperatures without the SQUID amplifier are approximately 12 K, which is the upper limit of the temperature axis in the figure.  As shown in Fig. 2, there is more than a factor of 10 improvement at the best frequency, and an improvement of more than a factor of 3 over 400 MHz of bandwidth.  This broadband improvement of the order of $\sqrt{10}$ corresponds to approximately a factor of 10 increase in speed in most microwave measurements making this a very useful amplifier for a large class of cryogenic microwave measurements.  Amplifiers with optimal operation frequencies at 1.7 GHz, 6.3 GHz, 6.8 GHz, 7 GHz, and 7.2 GHz were all measured, and all showed their best system noise temperature to be close to 1 K, or three added noise photons at 7 GHz.  Given that the double isolators at the input of the amplifiers can have as much as 1 dB of loss, this implies that the intrinsic amplifier noise at the optimal operating point is at least as low as 0.76 K, or 4.5 times the standard quantum limit (where we define the standard quantum limit to be one half the photon energy divided by the Boltzmann constant). \cite{caves1982}.
	
	We now discuss qualitatively the physical mechanism of the noise in these amplifiers and how we believe it could be improved in the future.  The noise of DC-SQUIDs has been studied in detail both theoretically and experimentally for lower frequency operation \cite{clarkebook,martinisandclarke}.  In the literature, it is found that the noise depends linearly on the temperature of the shunt resistors, which is in general higher than the physical temperature of the amplifier packaging due to Joule heating \cite{wellstood}.  It is also found that the mixed-down noise from the Josephson frequency is a very significant factor in the overall noise level of the SQUID.  We observe in all our SQUID amplifiers that the noise and gain have peaks and dips that correspond to certain values of the voltage on the device, hence to certain specific values of the Josephson frequency.  We speculate that these peaks and dips correspond to resonances from higher harmonics of the input resonator.  This theory is consistent with the fact that we see a larger number of harmonics in both the voltage and the in-band microwave characteristics in amplifiers with lower fundamental frequency input resonators than we do in the higher frequency amplifiers.
	
\begin{figure}
\includegraphics{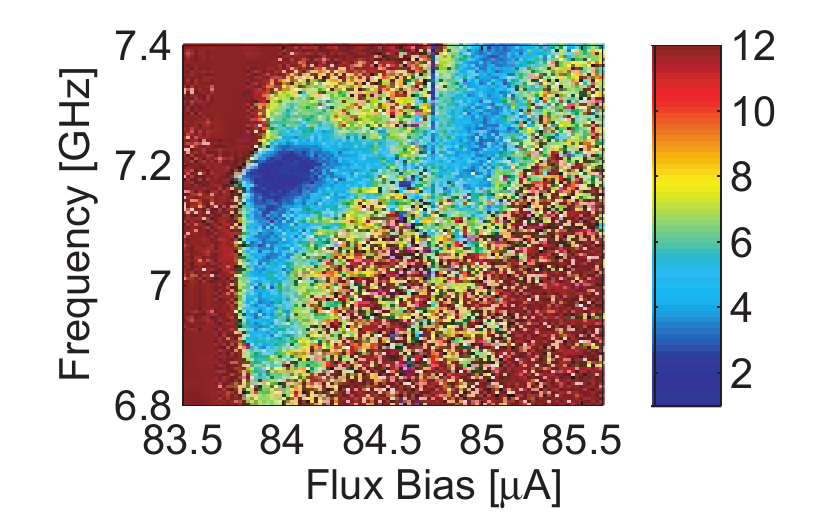}
\caption{System noise temperature in kelvin as a function of current through the flux bias coil and signal frequency for an amplifier with a 1.4 mm resonator.  Again, we see useful signal-to-noise improvement over several hundred MHz frequency range.  Also, note the strong dependence of optimal frequency, bandwidth, and minimum noise temperature on flux bias.  The current bias is fixed at 200 $\mu$A, which puts the SQUID in finite-voltage state.}
\end{figure}
		
	Figure 3 shows how noise temperature depends on both frequency and flux bias for a fixed current bias.  First, note that it is possible to change the optimal operating frequency of the amplifier by changing the bias.  We believe that this is due to the complex feedback dynamics of the SQUID, which change the effective reactance at the end of the input resonator, shifting the optimal signal frequency as well as the bandwidth \cite{mueck_feedback,spietz_squidamp2}.  Second, note that the bias point of the SQUID is absolutely critical in achieving optimal noise performance, and that care must be taken to use the amplifier at the correct bias point for a given application.  The narrow optimal point for the noise performance corresponds to a resonance in the voltage-current characteristics as well, both in the SQUID shown and in all of the other measured SQUIDs.  The exact nature of these resonances at higher harmonics and how they can be designed to improve noise performance will be explored in future work.	
	
	In conclusion, we have both shown useful noise characteristics in a practical microwave SQUID amplifier in the 4 GHz-8 GHz frequency range and have begun to learn what we need to in order to improve the performance in the future.  The bandwidth, operating frequency control, flatness of gain and noise, overall noise performance, and sensitivity to bias are all strong functions of how well controlled the microwave impedance environment is from the signal frequency up to at least the Josephson frequency of 20-30 GHz.  We plan to improve all of these figures of merit in future designs by measuring the scattering parameters of the lumped element SQUID in a calibrated way, and then using standard microwave engineering techniques to control the impedance environment as necessary to obtain the desired characteristics, in much the same way that an amplifier designer would with a transistor-based amplifier.  While the amplifier presented in this paper and previous work \cite{spietz_squidamp1,spietz_squidamp2} is already proving to be very useful for microwave quantum measurement experiments, these improvements will lead to an amplifier with even broader applicability by increasing the ease-of-use and ultimate noise performance considerably.

\bibliographystyle{apsrev4-1}


\end{document}